\documentclass[fleqn,usenatbib]{mnras}

\usepackage{newtxtext,newtxmath}

\usepackage[T1]{fontenc}
\usepackage{ae,aecompl}
\usepackage{xcolor}
\usepackage{ulem}
\usepackage{soul}

\usepackage{subfig}

\usepackage{graphicx}	
\usepackage{amsmath}	
\usepackage{verbatim}
\usepackage{bm}
\usepackage{multirow}

\usepackage{booktabs}

\newcommand{\change}[1]{{#1}} 

\graphicspath{{./plots/}}

\newcommand{\Msun}{M$_{\odot}$}

\newcommand{\gaia}{{\it Gaia}}

\defcitealias{cunningham19}{CU19}

\makeatletter
\DeclareRobustCommand{\I}{%
	\mbox{\check@mathfonts\fontsize\sf@size\z@\selectfont I}%
}
\DeclareRobustCommand{\V}{%
	\mbox{\check@mathfonts\fontsize\sf@size\z@\selectfont V}%
}
\makeatother

\title[Initial-final mass relation ]{Initial-final mass relation from white dwarfs within 40\,pc}

\author[T. Cunningham et. al]
{Tim Cunningham,$^{1,2}$\thanks{E-mail: tim.cunningham@cfa.harvard.edu }\thanks{NASA Hubble Fellow.}
Pier-Emmanuel Tremblay,$^{2}$ 
Mairi O'Brien$^{2}$ 
\\
$^{1}$Center for Astrophysics | Harvard \& Smithsonian, 60 Garden St., Cambridge, MA 02138, USA\\
$^{2}$Department of Physics, University of Warwick, Coventry, CV4 7AL, UK\\
}

\date{Accepted XXX. Received YYY; in original form ZZZ}

\pubyear{2023}

\begin{document}
\label{firstpage}
\pagerange{\pageref{firstpage}--\pageref{lastpage}}
\maketitle

\begin{abstract}
We present an initial-final mass relation derived from the spectroscopically-complete volume-limited 40\,pc sample of white dwarfs. The relation is modelled using population synthesis methods to derive an initial stellar population which can be fit to the observed mass distribution of white dwarfs. The population synthesis accounts for binary evolution, where higher-mass white dwarfs are more likely to be merger products than their lower-mass counterparts. Uncertainties are accounted from the initial mass function, stellar metallicity and age of the Galactic disc. We also consider biases induced by the spectral type of the white dwarf where pure-hydrogen atmosphere white dwarfs are likely to have more accurate masses, whilst the full white dwarf sample will have fewer biases arising from spectral evolution. We provide a four-piece segmented linear regression using Monte Carlo methods to sample the 1-$\sigma$ range of uncertainty on the initial stellar population. The derived initial-final mass relation provides a self-consistent determination of the progenitor mass for white dwarfs in the Solar neighbourhood which will be useful to study the local stellar formation history.

\end{abstract}




\section{Introduction}
\label{sec:intro}

The initial-final mass relation (IFMR) between the zero-age main-sequence and white dwarf masses provides a key diagnostic for  mass-loss at the end of stellar evolution. Main sequence stars with initial masses less than 8--10\,\Msun\ \change{(depending on metallicity)} are expected to evolve into white dwarfs \cite{iben1997} with more massive progenitors leading to neutron stars or black holes. The expected mass loss at the later stages of stellar evolution through the red giant branch (RGB), asymptotic giant branch (AGB), and post-AGB phase, is strongly mass dependent \citep{bloecker1995}. The amount of mass loss provides constraints on the physics of radiation, convection, chemical diffusion and mixing, nucleosynthesis, and angular momentum transport in the stellar envelope and interior. The mass loss is also dependent on stellar properties such as metallicity \citep{mcdonald2015} and rotation \citep{holzwarth2007,Cummings2019}.
AGB stars are important drivers of the evolution of galaxies, contributing to their integrated spectra \citep[\change{e.g.,}][]{Maraston2006}.

Mass loss rates due to stellar winds in RGB and AGB stars are typically observed in the range $10^{-8}$--$10^{-5}$\,M$_{\odot}/\rm{yr}$ 
\citep{iben1983,hofner2018}.
A more dominant form of mass loss in giant branch stars comes in the form of envelope ejection which \change{may produce} 
a planetary nebula. Typical planetary nebulae masses suggest a lower limit on the mass loss rate of $\sim10^{-5}$\,\Msun/yr \citep{renzini1981} during envelope ejection. In terms of total mass loss from the main sequence to white dwarf phase, empirical studies of the IFMR from white dwarfs in star clusters and wide binaries suggest that low- and intermediate-mass stars typically liberate between 20--80\% of their mass by the time they evolve into white dwarfs \citep{Weidemann1987,weidemann2000,catalan2008,kalirai2008,salaris2009,williams2009,dobbie2012,andrews2015,cummings2016,cummings2018,Barrientos2021,richer2021}. Recent work by \citet{marigo2020,Marigo2022} has suggested that at 1.5 to
2.25 initial masses, corresponding to white dwarf masses 0.6--0.7\,\Msun, the IFMR has a non-monotonic kink. Their proposed theoretical interpretation links the kink to thermally-pulsing AGB stars with a modest atmospheric carbon enrichment, caused by the third dredge-up, that is too low to trigger a powerful wind, prolonging the thermally-pulsing phase and allowing for carbon-oxygen core grow. At larger initial masses of 3--4\,\Msun, depending on metallicity, most theoretical IFMRs have a change in slope resulting from the second dredge-up which only occurs for higher
masses and reduces the AGB core mass \citep{marigo2007,meng2008,dominguez1999,cummings15,choi2016,Cummings2019}.

The mass-dependence of mass loss, coupled with the distribution of initial masses \change{together with metallicity} are the primary drivers of the resulting distribution of white dwarf masses. As a result it is feasible to compare an observed mass distribution of white dwarfs with an initial population to encapsulate the mass-dependent mass loss in the IMFR \citep{elbadry2018}. The mass distribution of single field white dwarfs is sharply peaked at $\approx$ 0.6\,\Msun, but a population of higher mass white dwarfs (0.8--1.33\,\Msun) is also well established \citep{bergeron92,koester2009-SPY,gianninas2011,kepler16,tremblay16,kilic2020,kilic2021}. The single-star progenitors for these high-mass white dwarfs are predicted to be main sequence stars in the mass range $\approx$ 3--9\,\Msun\ \citep[\change{e.g.,}][]{choi2016}.
However, it is also likely that a significant fraction of these stars are the products of binary mergers. Population synthesis studies \citep[\change{e.g.,}][]{temmink2020} predict that the fraction of merger products in the white dwarf mass range 0.8--1.33\,\Msun\ is likely to be $\approx$ 0.4, whereas at around 0.6\,\Msun\ the merger fraction is predicted to be closer to 0.2. \citet{Cheng20} estimate the fraction of double white dwarf mergers for $M_{\rm WD} > 0.8$\,\Msun\ at about 0.2 from \gaia\ observations.

In this study we derive an initial-final mass relation taking advantage of the \textit{Gaia} defined volume-limited 40\,pc white dwarf sample, with almost complete ($>$97\%) medium-resolution optical spectroscopic coverage \citep{OBrien2023}. Our approach is similar to that of \citet{elbadry2018} who adopted the volume-limited 100\,pc \gaia\ white dwarf sample with effective temperatures ($T_{\rm eff}$) above 10\,000\,K, except that here we do not use a temperature cut-off and the availability of spectroscopy allows for possibly more accurate white dwarf mass determinations. In Section\,\ref{sec:sample} we describe the volume-limited, spectroscopic sample of white dwarfs used as the final masses in the model. In Section\,\ref{sec:PopSyn} we describe the population synthesis model and explore the dependence of the IFMR on the initial parameters of the simulation. In Section\,\ref{sec:results} we present the final IFMR and the statistical uncertainty based on the uncertainties in the choice of initial parameters.
We also present the mass loss resulting from our population synthesis model and explore the implications for the theoretical understanding of mass loss at the end of stellar evolution and conclude in Section\,\ref{sec:conclusions}. 
 
\section{Sample}
\label{sec:sample}

\subsection{The 40pc white dwarf sample}
We use the volume-complete spectroscopic sample of white dwarfs within 40\,pc. A detailed discussion of the sample is given in \citet{limoges15,tremblay2020,mccleery2020,Gentile2021,OBrien2023} and O'Brien et al., submitted. This sample is based on the catalogue of white dwarf candidates from \citet{Gentile2021} drawn from \textit{Gaia} EDR3 \citep{Gaia2021}. Recent medium resolution ($R \approx 2000$) spectroscopic follow-up efforts \citep[][O'Brien et al., submitted]{limoges15,tremblay2020,OBrien2023} have now confirmed 1069 \gaia\ white dwarf candidates within 40\,pc out of the 1083 from \citet{Gentile2021}. We note that the white dwarf completeness of \gaia\ EDR3 is expected to be fairly high ($>$97\%) at 40\,pc based on the recovery rate of previously known white dwarfs \citep{hollands18b,mccleery2020}.

We rely on the photometric \gaia\ atmospheric parameters derived in \citet{Gentile2021} using pure-H \citep{tremblay11}, pure-He \citep{cukanovaite2021} and mixed H/He = 10$^{-5}$ in number \citep{tremblay14} model atmospheres and spectra. The grid of mixed model atmospheres is a best fit to the B-branch bifurcation in the \gaia\ Hertzsprung-Russell diagram \citep{Bergeron2019}, with hydrogen used as a proxy to represent both trace carbon and hydrogen \citep{Camisassa2023,Blouin2023}. For each white dwarf, we select one set of atmospheric parameters for the one chemical composition that best represents the spectral type and the  spectroscopic analyses from the literature, as described in table 2 of \citet{mccleery2020}.

\begin{figure}
	\centering
 	\subfloat{\includegraphics[width=1.\columnwidth]{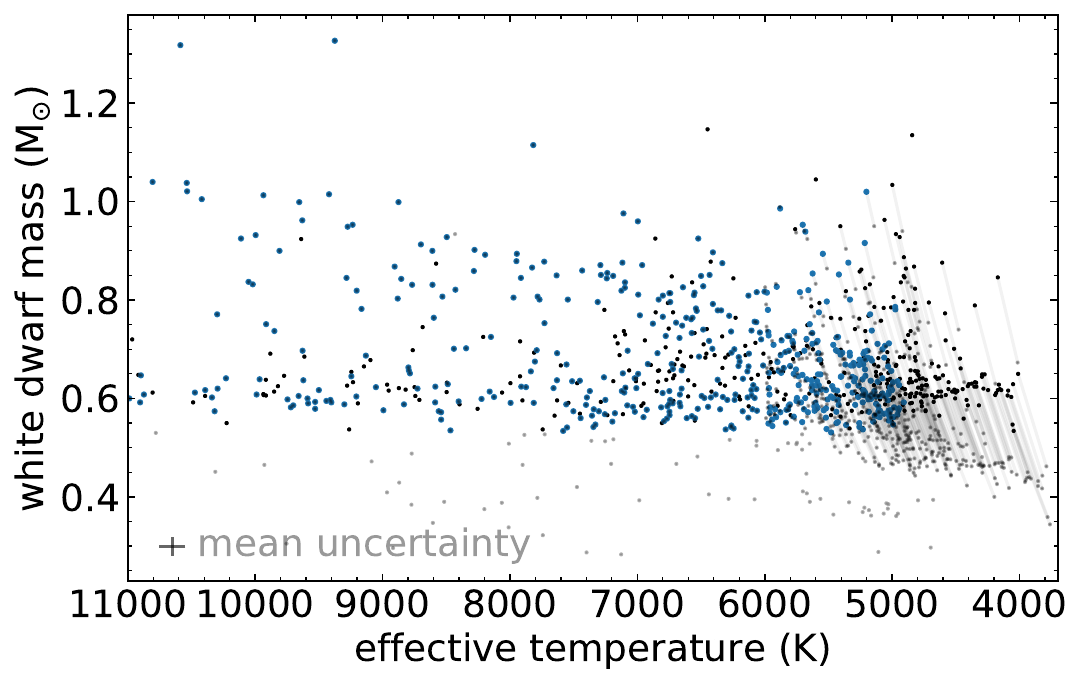}}
	\caption{Photometric mass and effective temperature distribution for the sample of 1069 \gaia\ white dwarfs within 40\,pc. For white dwarfs with masses greater than 0.53\,\Msun, spectral types DA and non-DA are shown in blue and black respectively. Those less massive are shown in grey. The mean uncertainty on the mass and effective temperature for the full sample is also indicated. \change{We also show the effect of the atmospheric parameter correction which is described in O'Brien et al., submitted.}}
	\label{fg:mass-teff}
\end{figure}

In Fig.\,\ref{fg:mass-teff} we show the mass and $T_{\rm eff}$ distribution of the full sample. We adopt a mass correction at low $T_{\rm eff}$ ($<$6000\,K) following a methodology similar to that discussed in \citet{cukanovaite2023} and the same as that presented in O'Brien et al., submitted\footnote{Table A1 from O'Brien et al., submitted, contains the updated parameters and has been made available at \url{https://cygnus.astro.warwick.ac.uk/phrtxn/}}. The correction ensures that the median mass of white dwarfs cooler than 6000\,K is the same as the median mass of the 40\,pc sample at larger temperatures, as is expected from a population of non-interacting white dwarfs cooling at constant mass \citep{tremblay16}. This addresses the low-mass problem arising from missing physics in the atmospheric models, where the issue is observed when using both \gaia\ and Pan-STARRS photometry, as well as independent models and fitting methods \citep{hollands18b,Bergeron2019,Blouin19,tremblay2020,mccleery2020,OBrien2023}.

\begin{figure}
	\centering
	  \subfloat{\includegraphics[width=1.\columnwidth]{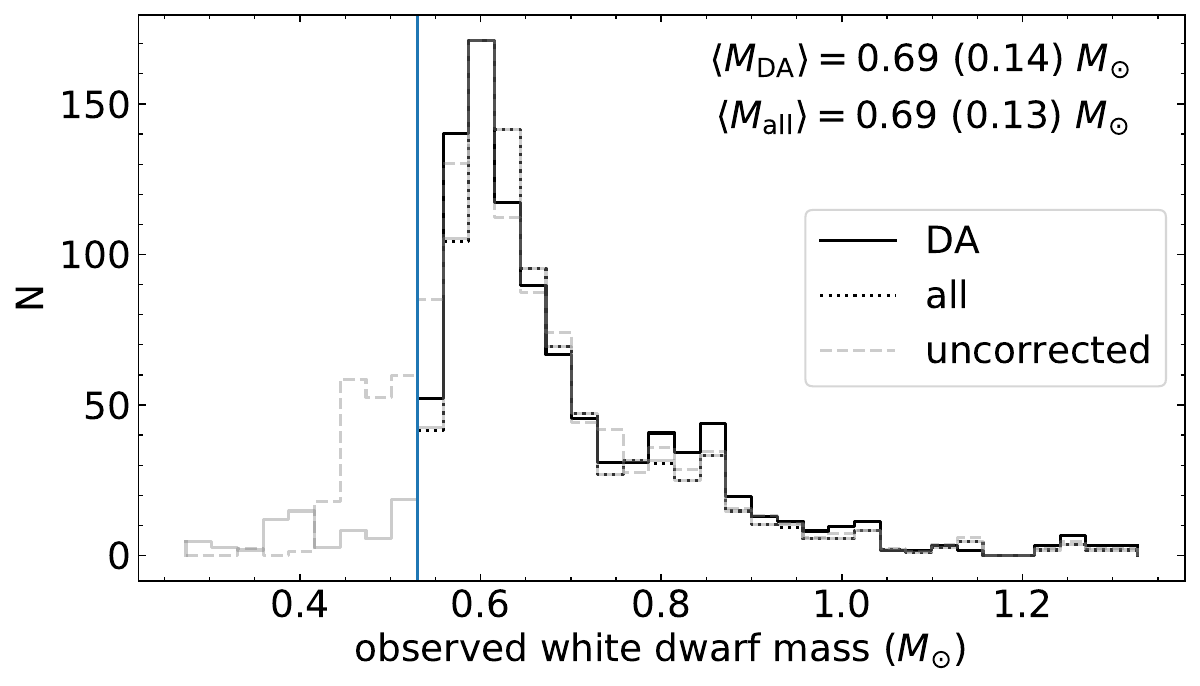}}
	\caption{Photometric mass distribution for the sample of 1069 \gaia\ white dwarfs within 40\,pc. The full sample (dotted line) is compared to the subsample of 655 DA white dwarfs (solid line). The blue horizontal line represents our lower mass cutoff to clean off the sample from double degenerate candidates (gray mass distribution). The mass cut leaves 963 in the full sample, and 590 in the DA-only sample. The mean mass and standard deviation for the samples above the mass cut are also shown. \change{We also show the mass distribution before the atmospheric parameter correction (described in O'Brien et al., submitted) was applied.}}
	\label{fg:mass-dist-DA-all-low}
\end{figure}

In Fig.\,\ref{fg:mass-dist-DA-all-low} we show the mass distribution for the full sample and for the hydrogen-dominated atmosphere white dwarfs (DAs) only. We find no significant difference between the mean and standard deviation values of the two distributions. The blue vertical line indicates a mass cut of 0.53\,M$_{\odot}$ which we adopt to remove white dwarfs that may result from binary evolution and double degenerate candidates, which likely have incorrect \textit{Gaia} masses because of the assumption of a single star in the fitting procedure. This cut leaves a sample of 963 confirmed white dwarfs. A similar cut of 0.54\,\Msun\ was adopted by \citet{cukanovaite2023} to derive to local stellar formation history and broadly corresponds to the intersection of the distributions of single white dwarfs peaking at 0.6\,\Msun\ and the double degenerates below this. The majority of double degenerate candidates within 40\,pc are unconfirmed while the minimum mass of a white dwarf than can be formed from single star evolution within the lifetime of the Galactic disk has an uncertainty of several percent \change{and depends on the metallicity of the progenitor}  \citep{kalirai12,cummings2018,elbadry2018,marigo2020}. Therefore, the adopted sharp cutoff was judged to be the best compromise given observational constraints. In Section\,\ref{sec:results} we fit the IFMR using both the full and DA-only samples.

The advantage of the DA-only sample is that the \textit{Gaia}-derived masses are more reliable than for the non-DA white dwarfs, where traces of hydrogen, carbon or metals in a helium-dominated atmosphere can have a significant effect on its mass determination \citep{Bergeron2019,Blouin2023,Camisassa2023}. The DA-only sample also greatly reduces the influence of the ad-hoc correction for the low-mass problem imposed below 6000\,K, since most white dwarfs below 5000\,K are of DC spectral type due to the lack of observable hydrogen or helium lines. On the other hand, the full sample alleviates the biases induced by spectral evolution in which some DA white dwarfs below 18\,000\,K evolve into non-DA spectral types due to convective mixing \cite[see, e.g.,][]{cunningham20,Lopez2022,ourique2020}. This may introduce a small bias if spectral evolution is mass dependent.

\section{Population Synthesis Techniques}
\label{sec:PopSyn}

\begin{table*}
	\centering
        \caption{Parameters sampled using Monte Carlo methods to quantify the astrophysical scatter in the IFMR.}
        \begin{tabular}{lll}
                \hline            
                \vspace{3pt}
Parameter & Values & Distribution\\
\hline \vspace{6pt}
IMF slope & $\mu=2.35, \sigma=0.1$ & Gaussian \\ 
\vspace{6pt}
Metallicity & $\mu=0.02, \sigma=0.25$ & Gaussian \\ 
\vspace{6pt}
Age of the Galactic disk & $\mu=10\,\mathrm{Gyr}, \sigma=0.7\,\mathrm{Gyr}$ & Gaussian ($<$11\,Gyr) \\ 
\vspace{6pt}
Spectral types & 1) DA, or 2) DA+non-DA & Binary\\ 
\vspace{2pt}
Merger products & 1) products removed, or 2) not & Binary\\ 
                \hline
        \end{tabular}
        \label{tb:MC-parameter-space}\\
\end{table*}

We use the technique of population synthesis to initialise a population of main sequence stars. \change{We initially create a population using the following assumptions:} 
1) constant stellar formation rate, which has recently been shown to be appropriate for the 40\,pc sample \citep{cukanovaite2023}, 2) \cite{salpeter1955} initial mass function (IMF), 3) main sequence lifetimes determined for Solar metallicity from the models of \cite{hurley2000} with a 4) Galactic disk age of 10\,Gyr \citep{cukanovaite2023}, \change{although we explore the uncertainty on these assumptions later.}

\change{For a given set of assumptions about the main sequence population (e.g, metallicity, IMF slope, star formation rate), the}
IFMR provides a unique transformation from the initial synthetic population to the observed mass distribution within 40\,pc. Typically this relation is assumed to be monotonic, although evidence has been presented for non-monotonicity \citep{marigo2007,kalirai2014,marigo2020}. We rely initially on the quantile-quantile relation between the two mass distributions: 1) the stars in the synthetic initial population which have total ages in excess of their main sequence lifetime and 2) the observed distribution of white dwarfs within 40\,pc. The quantile-quantile relation provides a direct mapping from one to the other, assuming only a monotonic relation, as in almost all previous studies, with no functional form. This approach allows us to probe the parameter space of the population synthesis to determine the dependency of the IFMR on the choice of initial parameters.

In order to probe the scatter in the IFMR from astrophysical parameters we derive a unique IFMR for each initial population drawn from the range of parameters described in this section (see Table\,\ref{tb:MC-parameter-space}) and plotted in Figs.\,\ref{fg:quantile-param-vary}\,\&\,\ref{fg:quantile-param-vary-DA-non-DA-merger}. We use fifty evenly spaced quantiles which ensures the observed population still has a statistically significant number of members in each bin defined by the distance between each quantile. We note that with arbitrarily large numbers of quantiles it is always possible to recreate exactly the observed white dwarf distribution.
Our quantile approach recovers a ``knee'' at initial masses of $\approx$3.5\,\Msun\ that was noted by \citet{elbadry2018} (most clearly seen in Fig.\,\ref{fg:quantile-param-vary-DA-non-DA-merger}), although we note it is less pronounced compared to the IFMR of \citet{elbadry2018}.
Whilst it is possible to fit the observed distribution exactly for one configuration of the initial population, we need to provide a function that accounts for the intrinsic scatter in a statistical fashion. 

Our approach is to draw an initial synthetic population and observed white dwarf distribution using Monte Carlo (MC) methods and the range of parameters detailed in Table\,\ref{tb:MC-parameter-space}. The IFMR is derived by computing the quantile-quantile relation between the two distributions using evenly-spaced quantiles, which is equivalent to using evenly-populated bins. This process is repeated for $N$ = 10\,000 draws to build up a statistically-robust estimate of the scatter on the IFMR due to uncertainties in the IMF, initial stellar metallicity, Galactic disk age, and white dwarf sample biases. We will describe each of these uncertainties in the following sections.

\subsection*{Initial mass function}

We adopt an IMF which scales as a power law $dN/dm\propto m^{-\alpha}$ where the \citet{salpeter1955} IMF has a slope of $\alpha=2.35$. To account for uncertainties induced by the IMF slope, we draw slopes ($\alpha$) from a normal distribution with mean $\alpha=2.35$ and standard deviation of 0.1 (e.g., \citealt{elbadry2018,weisz2015}). The top-left panel of Fig.\,\ref{fg:quantile-param-vary} shows the influence of this parameter range on the IFMR. The inset shows the distribution of MC samples of the IMF slopes. There are many prescriptions of the IMF available in the literature (e.g., \citealt{kroupa2001,chabrier2003,miller1979,maschberger2013,ferrini1990}). The majority of the more recent IMF prescriptions than the Salpeter IMF favour a shallower slope at low initial mass ($M_{\rm initial}\lessapprox 1.0\,$M$_{\odot}$). The main sequence lifetime for the majority of these stars is longer than the age of the Galactic disc and so this parameter space is of small significance for the local population of white dwarfs. 

\subsection*{Main-sequence lifetime/metallicity}
We compute main sequence lifetimes from the models of \citet{hurley2000} in order to establish whether a synthetic star has reached the white dwarf phase at current time. We draw from a distribution of metallicity, adopting the mean value of Solar metallicity at $Z=0.02$ (\citealt{vagnozzi2019} found Z$_{\odot}=0.0196 \pm 0.0014$). The distribution of metallicity in the Solar neighborhood for G-dwarfs was found to have a standard deviation of 0.2\,dex from Solar (see Figure 3 of \citealt{haywood2001}). More recently, \citet{buder2019} found a mean and standard deviation of metallicity of $-0.0427 \pm 0.0019$\,dex and $0.2461 \pm 0.0009$\,dex, respectively. For our population synthesis we draw from a normal distribution in metallicity with \change{the mean value set to Solar}
and a standard deviation of 0.25\,dex. The impact of this distribution on the IFMR is shown in the top-right panel of Fig.\,\ref{fg:quantile-param-vary}.

\subsection*{Galactic disk age}
We draw from a distribution of Galactic disk ages with a mean of 10\,Gyr and standard deviation of 0.7\,Gyr and maximum allowed value of 11\,Gyr. 
\change{This assumption may exclude halo stars although they only accounts for 1-2\% of the local white dwarf population \citep[see, e.g.,][]{mccleery2020}.}
The standard deviation illustrates typical systematics in the white dwarf cooling ages predicted by different groups \citep{Salaris2013,Camisassa2016,camisassa2019,bedard2020}. The impact of this range of maximum ages is shown in the bottom right panel of Fig.\,\ref{fg:quantile-param-vary}. 

\subsection*{Stellar mergers and atmospheric composition}

\change{We alter the synthetic mass distribution}
to remove likely merger products which do not fit in our single-star IFMR. The population synthesis binary models of \citet{temmink2020} predict that $\approx$ 40\% of single white dwarfs with masses above $\approx$ 0.8\,M$_{\odot}$ are likely to be the product of WD+WD, WD+MS or MS+MS mergers in their ``default'' model which we use in this study. We set out to produce a single star IFMR and so stochastically remove white dwarfs from the observed mass distribution based on the probability of being a merger product.
The top-panel of Fig.\,\ref{fg:mass-dist-merger-removed} shows the merger fraction from \citet{temmink2020}.
The distribution of DAs before and after this merger removal process can be seen in Fig.\,\ref{fg:mass-dist-merger-removed}. As expected the latter distribution is more biased towards lower mass white dwarfs.

In the bottom panel of Fig.\,\ref{fg:quantile-param-vary-DA-non-DA-merger} we show the IFMR derived from the mean synthetic initial population with the two observational samples. In the top panel we show the same, but for the DA-only and full samples. Broadly, we find that this choice of observed distributions does not make a large impact on the IFMR. For the final IFMR presented in this work we draw Monte Carlo samples from the DA-only and full samples, in both cases with merger-product removal.
We conclude the fraction of mergers not to be a dominant source of uncertainty in our IFMR.

\begin{figure*}
	\centering
	\subfloat{\includegraphics[width=0.89\textwidth]{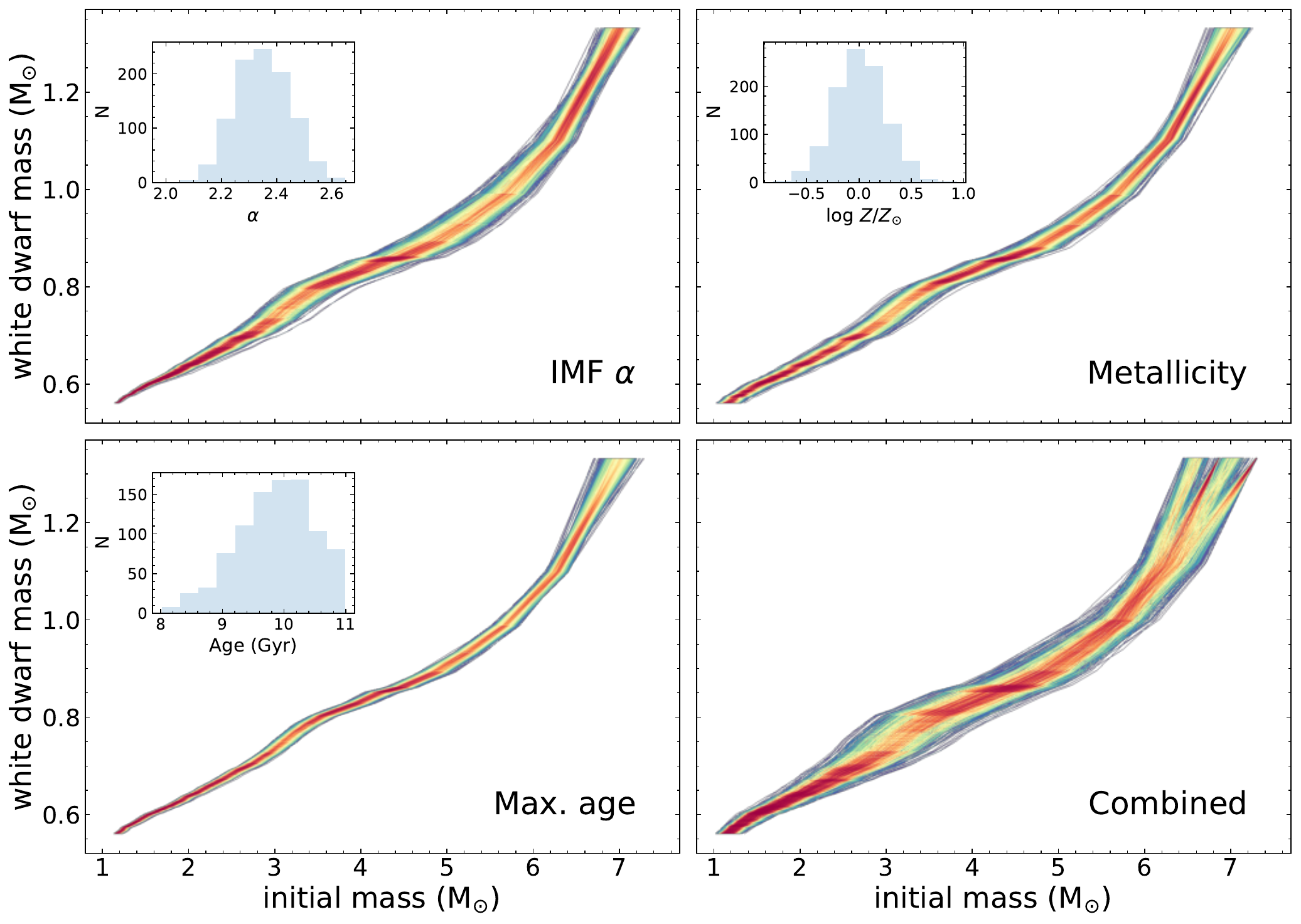}}
	\caption{Varying parameters in the population synthesis. We show the quantile-quantile IFMR derived using fifty evenly-spaced quantiles to map monotonically from the synthetic initial mass distribution to the observed 40\,pc white dwarf mass distribution. The color represents the density of lines. Here we vary three parameters in the initial population -- \textit{top-left}: IMF slope ($\alpha = 2.35\pm 0.1$), \textit{top-right}: metallicity ($Z/Z_{\odot}=0.00\pm0.25$), and \textit{bottom-left}: age of the Galactic disc ($A=10.0\pm0.7$\,Gyr). \textit{Bottom-right}: we show the combined Monte Carlo sampling of the three parameters in the other panels, where we also include the uncertainty on the observed distribution as described in Fig.\,\ref{fg:quantile-param-vary-DA-non-DA-merger}. }
	\label{fg:quantile-param-vary}
\end{figure*}

\begin{figure}
	\centering
	\subfloat{\includegraphics[width=0.80\columnwidth]{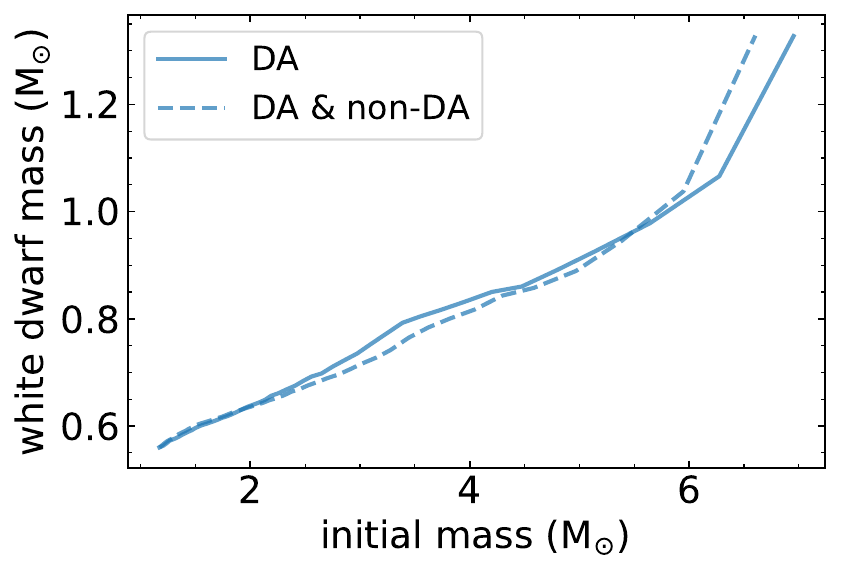}}\\
	\subfloat{\includegraphics[width=0.80\columnwidth]{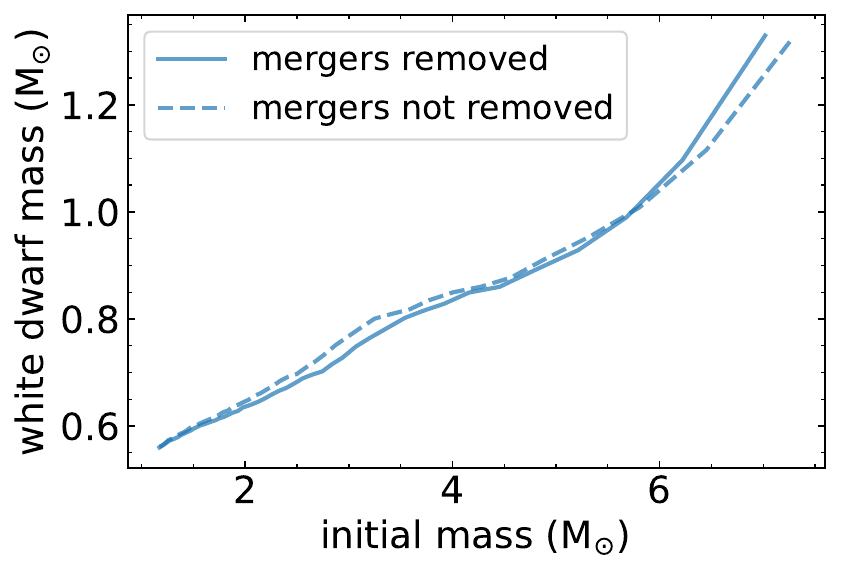}}
	\caption{Similar to Fig.\,\ref{fg:quantile-param-vary} for the DA-only and full samples (top) and either with or without the stochastic removal of merger products from the observed distribution (bottom). Here we only take the mean initial population ($\alpha=2.35$, $Z/Z_{\odot}=0.00$, and $\rm{Age}=10$\,Gyr) to isolate the effect of using the different observed distributions.}
	\label{fg:quantile-param-vary-DA-non-DA-merger}
\end{figure}

\begin{figure}
	\centering
	\subfloat{\includegraphics[width=.9\columnwidth]{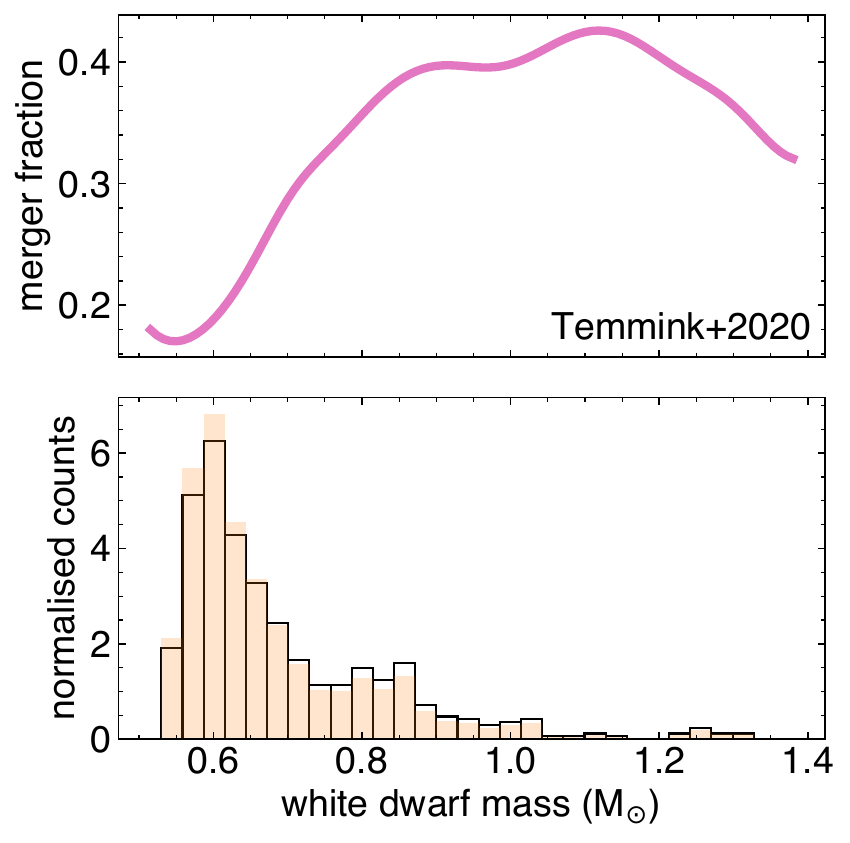}}
	\caption{{\it Top panel:} The theoretical probabilistic merger fraction from \citet{temmink2020} as a function of white dwarf mass. {\it Bottom panel:} Mass distribution of the DAs in the 40pc sample (black) and mass distribution after the removal of likely merger products (orange).}
	\label{fg:mass-dist-merger-removed}
\end{figure}

\begin{figure}
	\centering
	\subfloat{\includegraphics[width=0.80\columnwidth]{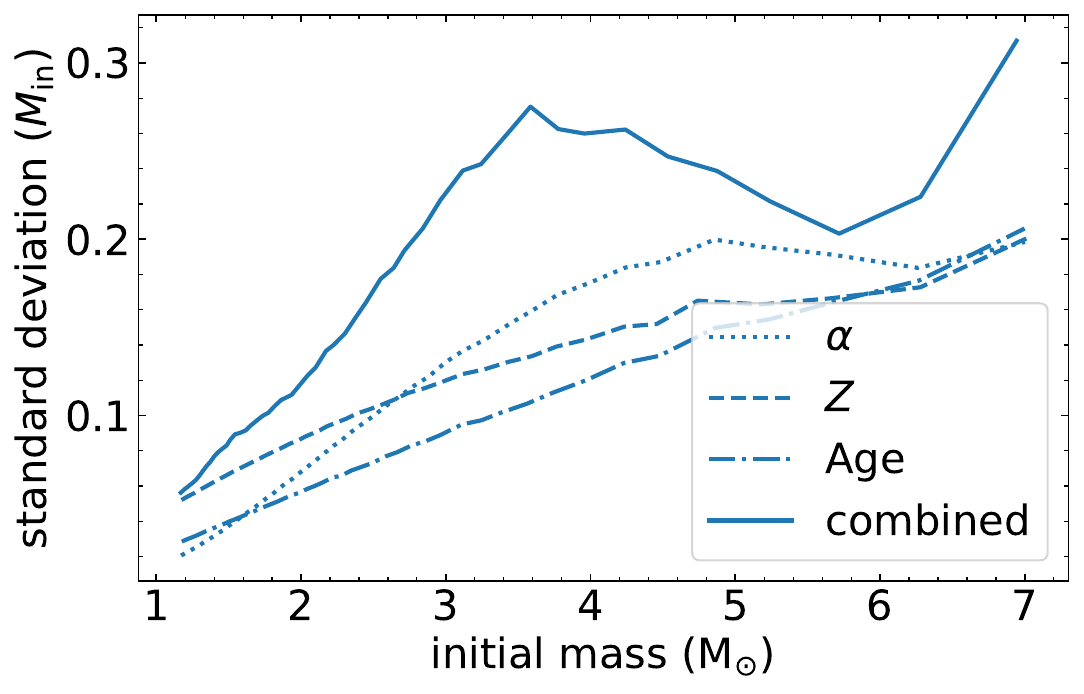}} \hfill
	\caption{Standard deviation in the initial mass for the parameter study shown in Fig.\,\ref{fg:quantile-param-vary} which includes a Monte Carlo sampling of initial mass function slope ($\alpha$), progenitor metallicity ($Z$), age of the Galactic disk and all parameters combined (including DA/non-DA and merger products removed or not).}
	\label{fg:quantile-param-vary-stdv}
\end{figure}

In Fig.\,\ref{fg:quantile-param-vary-stdv} we show as a function of initial mass the standard deviation in initial mass for the IFMRs shown in the first three panels in Fig.\,\ref{fg:quantile-param-vary} -- IMF slope, metallicity, and Galactic disk age. We find that for the ranges considered, each parameter provides about a 3\% standard deviation on the initial mass values. The combination of all three effects, as well as the DA/full sample and merger removal, produces a 5--8\% standard deviation in the IFMR. 

\subsection*{High mass progenitors}
Constraining the IFMR at the high-mass ($M_{\rm WD}>1.0$ M$_{\odot}$) end presents a challenge with the sample used in this work due to the low number of high-mass white dwarfs within 40\,pc (only 3\% of the sample have masses in excess of 1.0\,\Msun). 
However, a key source of uncertainty in the IFMR is the progenitor mass for the highest mass white dwarfs measured in our sample ($\approx$ 1.3\,$M_{\odot}$), 
where the high-mass slope of the IFMR is sensitive to the largest mass of each distribution.
For the observed distribution this is determined by the largest mass white dwarf in 40\,pc. For the synthetic population this is harder to define since we do not a priori know the mass of the progenitor to a white dwarf at 1.3\,M$_{\odot}$. 
We tackle this issue by calibrating the highest-mass bin of our IFMR to that of \citet{cummings2018} who derived the IFMR from open and globular clusters. 
\change{The open and globular clusters used in \citet{cummings2018} may have different metalicities to the main-sequence stars in the Solar neighbourhood. However, the authors found no evidence of metal-dependence in their IFMR in the range 
$-0.15 < $ [Fe/H] $< +0.15$.
Nonetheless, an}
alternative approach would be to set the progenitor-mass of a Chandrasekhar mass white dwarf (e.g., \citealt{elbadry2018}). We find that this would make no discernible difference on our IFMR as we recover the \citet{elbadry2018} IFMR at high masses, despite being calibrated against \citet{cummings2018}.

\subsection*{Low mass progenitors}

We fit our IFMR by comparing the quantiles of the synthetic initial distribution and the observed white dwarf mass distribution. The observed distribution has a mean measurement uncertainty of 0.02\,\Msun\ \citep{Gentile2021}.
Due to this uncertainty in the measured white dwarf parameters, we expect the lower-mass wing of the mass distribution ($0.53<M_{\rm}/$M$_{\odot}<0.60$) to be a Gaussian tail to a sharper \textit{true} peak near 0.56--0.6\,\Msun. 
We thus derive the lowest-mass point of the IFMR at white dwarf masses of 0.56\,\Msun, accounting for the fact that 6\% of the sample have masses below that (i.e., in the range 0.53--0.56\,\Msun). For white dwarf masses below 0.56\,\Msun\ we assume the IFMR to be a linear extrapolation of the best-fit slope in the $\approx$0.56--0.65\,\Msun\ white dwarf mass range. 

\subsection*{Completeness}

\begin{figure}
	\centering
	\subfloat{\includegraphics[width=0.89\columnwidth]{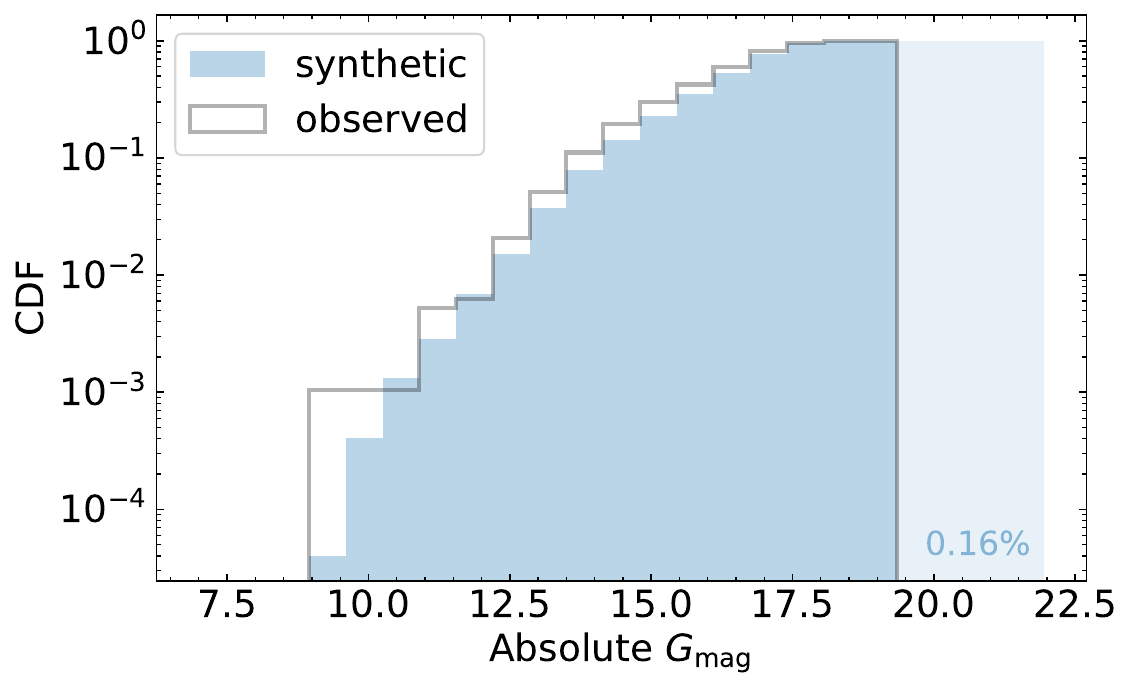}}
	\caption{Cumulative distribution of synthetic apparent \textit{Gaia} magnitudes (shaded blue), compared to the observed distribution (black). Apparent magnitudes are computed using the IFMR derived in this study and cooling models of \citet{bedard2020}.}
	\label{fg:Gaia-Gmag}
\end{figure}

Close to the \gaia\ $G$ magnitude limit of 20--21, white dwarf identification becomes increasingly difficult \citep[see, .e.g.,][]{Gentile2021}.
In Fig.\,\ref{fg:Gaia-Gmag} we show the cumulative distribution of apparent $G$ magnitudes for the synthetic population of white dwarfs. Apparent magnitudes were computed using the cooling models of \citet{bedard2020} and synthetic colours of \citet{tremblay11,kowalski2006,cukanovaite2021}.
The distribution was made assuming a uniform density distribution of white dwarfs within the 40\,pc volume. We find that $<0.1$\% of the sample have $G>20$\,mag and $<0.2$\% have $G>19.35$\,mag, the faintest observed magnitude of a white dwarf in the 40 pc sample, which we adopt as a heuristic limiting magnitude. 
We conclude that the 40\,pc white dwarf sample is broadly complete and the number of missing fainter white dwarfs is likely to be small enough ($\approx$ 2) not to significantly influence the derived IFMR in this work. 

It is estimated that $\approx$ 3\% of the white dwarfs within 40\,pc are not recovered by the \gaia\ EDR3 selection of \citet{Gentile2021}, for the most part due to an unresolved main-sequence companion \citep{mccleery2020,OBrien2023}. Since we aim at deriving the IMFR representing single star evolution, we make no attempt to add back these objects to the sample.

\subsection*{Intrinsic scatter}

While we account for the metallicity dependence of the main sequence lifetime encapsulated in the analytic model of \citet{hurley2000}, it does not include the effect of metallicity on mass loss in the AGB phase. We also neglect the effect of stellar rotation where enhanced rotation during main sequence may \citep{friend1986,holzwarth2007} or may not \citep{owocki1996,glatzel1998} lead to enhanced mass loss and altered anisotropies. \citet{Cummings2019} find that rotational mixing with convective core-overshoot in main sequence stars, which creates more massive cores and extends the star's lifetime, is necessary to explain the observed IFMR at higher masses ($M_{\rm initial} > 3$\,\Msun).
The magnetic field strength of the star may also play a role in the mass loss \citep{quentin2018}. Theoretical models have predicted that in massive stars magnetic fields lead to lower rates of mass loss \citep{keszthelyi2019}. 

In our population synthesis model we do not account for rotation, AGB metallicity or magnetic fields. 
We note that IMFR studies using white dwarfs in wide binaries or clusters are more suitable to study the intrinsic scatter in the IMFR, because the total stellar age can then be constrained from a wide companion or the cluster main-sequence turn-off, respectively \citep{kalirai2008,catalan2008-wd-ms-binary,andrews2015,cummings2018,Barrientos2021}. In our approach using single white dwarfs within 40\,pc, it is impossible to directly constrain the scatter in the IFMR, although it is possible to determine the median IFMR to a high precision which is the main purpose of this work.

\section{Results and discussion}
\label{sec:results} 

The quantile-quantile approach implicitly assumes monotonicity in the IFMR, but does not assume any function. However, extracting a function from the quantile result is still desirable. We find that below $M_{\rm initial} \approx$ 6\,M$_{\odot}$ the distribution is insensitive to the choice of quantiles (bins). In this range, the choice of a piecewise linear fit which has been employed in  both previous observational and theoretical studies (e.g. \citealt{elbadry2018,cummings2018,Cummings2019})
is qualitatively appropriate (see Fig.\,\ref{fg:quantile-param-vary}). We identify a prominent breakpoint at $M_{\rm initial} \approx$ 3.5\,M$_{\odot}$ which \citet{cummings15} and \citet{elbadry2018} point out is predicted by stellar evolution models due to the helium flash and second dredge-up \citep{dominguez1999,marigo2007,choi2016}.

\begin{table}
	\centering
        \caption{Mean values of the breakpoints in the segmented linear fit shown in Fig.\,\ref{fg:ifmr-1sigma-compare} with 65\% confidence interval estimated using a Monte Carlo sampling of 1-sigma uncertainties on the quantile IFMRs shown in the bottom right panel of Fig.\,\ref{fg:quantile-param-vary}. The distributions underlying the 1-$\sigma$ uncertainties shown here can be seen in Fig.\,\ref{fg:pwlf-cornerplot}.}
        \begin{tabular}{ll}
                \hline            
                \vspace{5pt}
$M_{\rm initial}$ & $M_{\rm WD}$ \\
\hline \vspace{6pt}
$1.09_{-0.02}^{+0.02}$ & ${0.561}_{-0.002}^{+0.002  }$ \\ 
\vspace{6pt}
$2.65_{-0.19}^{+0.19}$ & ${0.70}_{-0.02}^{+0.02}$ \\ 
\vspace{6pt}
$3.42_{-0.21}^{+0.20}$ & ${0.79}_{-0.02}^{+0.02}$ \\ 
\vspace{6pt}
$5.06_{-0.27}^{+0.26}$ & ${0.91}_{-0.03}^{+0.02}$ \\ 
\vspace{2pt}
$7.44_{-0.27}^{+0.24}$ & ${1.30}_{-0.05}^{+0.05}$ \\ 
                \hline
        \end{tabular}
        \label{tb:MC-pwlf-breakpoints}\\
\end{table}

To provide a function that accounts for the uncertainties in a statistical fashion we turn to a segmented linear regression as has been employed in many previous studies. We find that a 4-segment linear fit provides a reasonable description to our observational sample. The 4-piece fit was employed by \citet{elbadry2018} in order to capture the ``knee'' at $M_{\rm initial} \approx$ 3.5\,\Msun, which we also recover from our quantile IFMRs (most clearly seen in Fig.\,\ref{fg:quantile-param-vary-DA-non-DA-merger}). The same figure implies that a steepening may be appropriate at high masses ($>$6.0\,\Msun). However, this is based on just one bin of data so we do not include a fifth segment to capture this feature. Instead we draw Monte Carlo samples from the ``combined'' quantile IFMRs (bottom-right panel Fig.\,\ref{fg:quantile-param-vary}) taking the mean and standard deviation of the distributions in the bins defined as the separation between each quantile. We draw a random sample of initial and final masses from this distribution and perform a 4-piece segmented linear regression using the the \texttt{Python} package \texttt{pwlf} \citep{pwlf}.

\begin{figure}
	\centering
	\subfloat{\includegraphics[width=0.89\columnwidth]{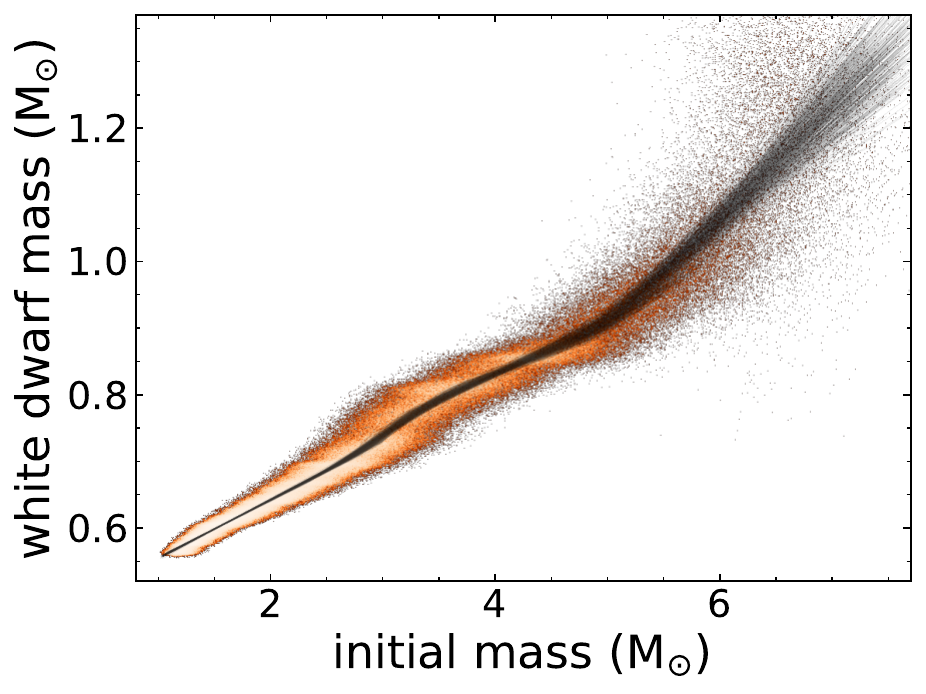}}
	\caption{Piecewise linear fits (black lines) to sampled quantiles from Monte Carlo combined simulated IFMR (bottom-right panel of Fig.\,\ref{fg:quantile-param-vary}). The fit is performed 500 times using the bootstrap method in order to estimate the uncertainty on the derived parameters. The distribution of best-fit break point locations is shown in the Appendix Fig.\ref{fg:pwlf-cornerplot}.}
	\label{fg:quantile-pwlf-MC-fits}
\end{figure}

Fig.\,\ref{fg:quantile-pwlf-MC-fits} shows the Monte Carlo samples and the best-fit segmented linear regression to each draw. As expected we find the largest uncertainty in the IFMR slope to be at the high-mass end, an effect which is reflected in most IFMR prescriptions in the literature. This is mostly due to the low numbers of observed high-mass white dwarfs and corresponding high-mass progenitors. In the Appendix Fig.\,\ref{fg:pwlf-cornerplot} we show the best-fit locations of the breakpoints for all of the black lines shown in Fig.\,\ref{fg:quantile-pwlf-MC-fits} along with the mean and standard deviations of the initial mass and white dwarf mass at each breakpoint. The mean breakpoint locations and the 1-$\sigma$ uncertainties are given in Table\,\ref{tb:MC-pwlf-breakpoints}.

\begin{figure}
	\centering
    \subfloat{\includegraphics[width=1.\columnwidth]{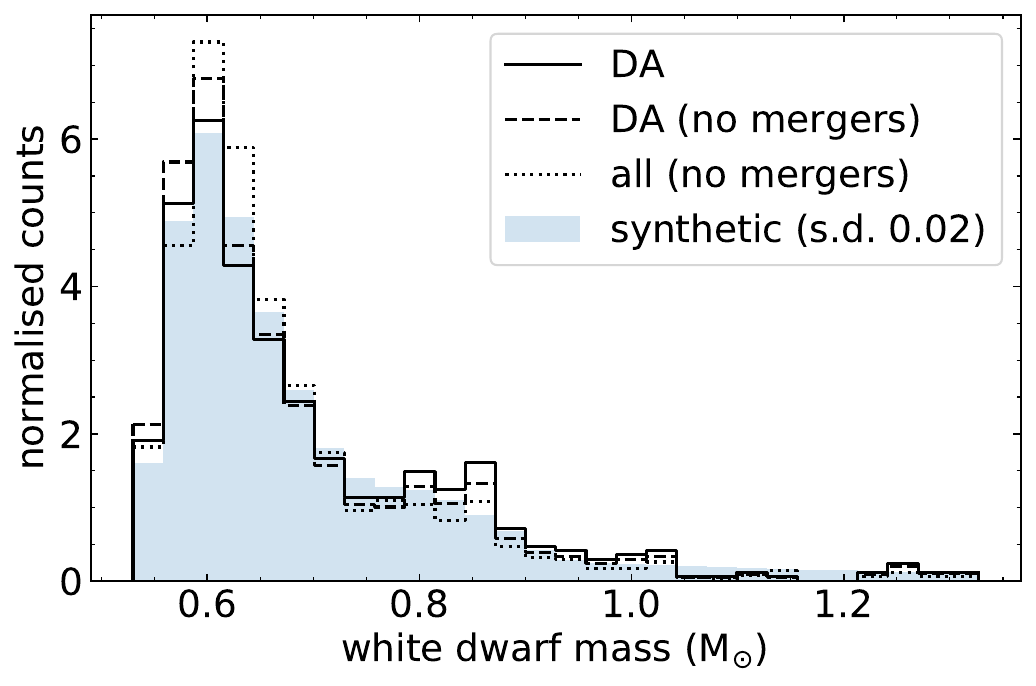}}
	\caption{Observed mass distribution of 40\,pc white dwarfs. We show the distribution for DA white dwarfs, the same distribution weighted by the merger fraction \citep{temmink2020}, and the merger-weighted distribution of all white dwarfs (DA and non-DA). In blue is the synthetic population of main sequence stars put through the IFMR developed in this work. The elements of the synthetic distribution have been subjected to a 1-$\sigma$ \gaia\ photometric mass uncertainty of 0.02\,M$_{\odot}$. 
 }
	\label{fg:final-hist-mass-dist}
\end{figure}

In Fig.\,\ref{fg:final-hist-mass-dist} we show the synthetic distribution of white dwarf masses resulting from putting the mean initial stellar population through the 
IFMR
derived in this study (see Table\,\ref{tb:MC-pwlf-breakpoints}). We give our synthetic white dwarfs a mass uncertainty drawn from a normal distribution with standard deviation 0.02\,\Msun, corresponding to the median uncertainty for the 40\,pc sample \citep{Gentile2021}. We do not include this source of uncertainty in the best-fit IMFR breakpoint location in Table\,\ref{tb:MC-pwlf-breakpoints}, but we do apply this additional uncertainty 
in the derivation of the synthetic white dwarf masses shown in Fig.\,\ref{fg:final-hist-mass-dist}. 

\begin{figure}
	\centering
	\subfloat{\includegraphics[width=0.89\columnwidth]{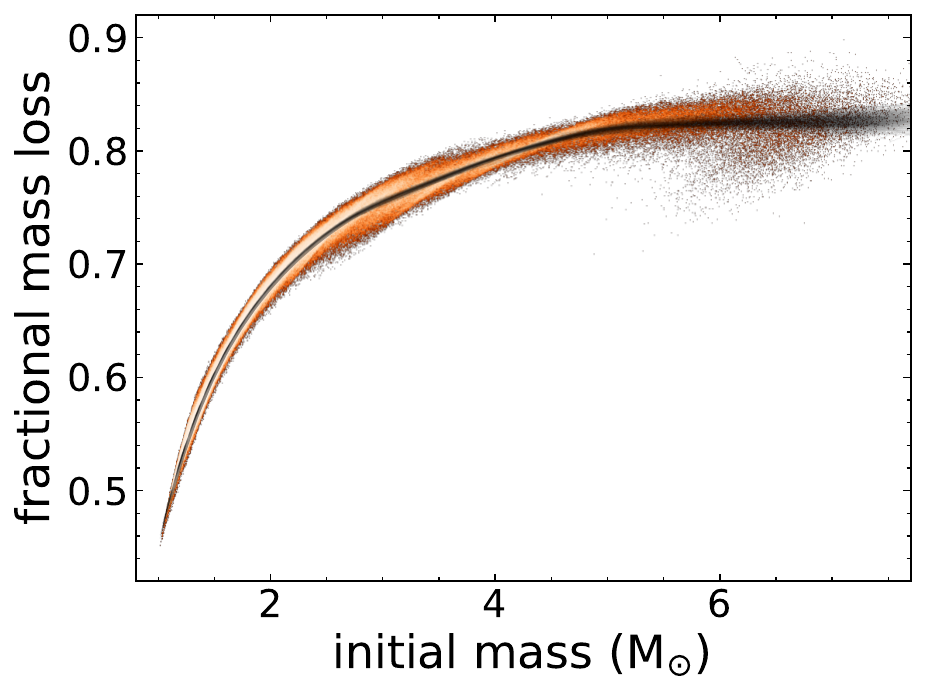}}
	\caption{Fractional mass loss based on the IFMR from Monte Carlo sampling shown in Fig.\,\ref{fg:quantile-pwlf-MC-fits}.}
	\label{fg:quantile-pwlf-MC-fits-mass-loss}
\end{figure}

Fig.\,\ref{fg:quantile-pwlf-MC-fits-mass-loss} shows the predicted mass loss from our Monte Carlo sampling of the IFMR accounting for the three dominant forms of astrophysical scatter discussed in Section\,\ref{sec:PopSyn}. Our model finds that the progenitors of the white dwarfs within 40\,pc are likely to have liberated between 45--85\% of their initial mass. 

\begin{figure}
	\centering
	\subfloat{\includegraphics[width=1.\columnwidth]{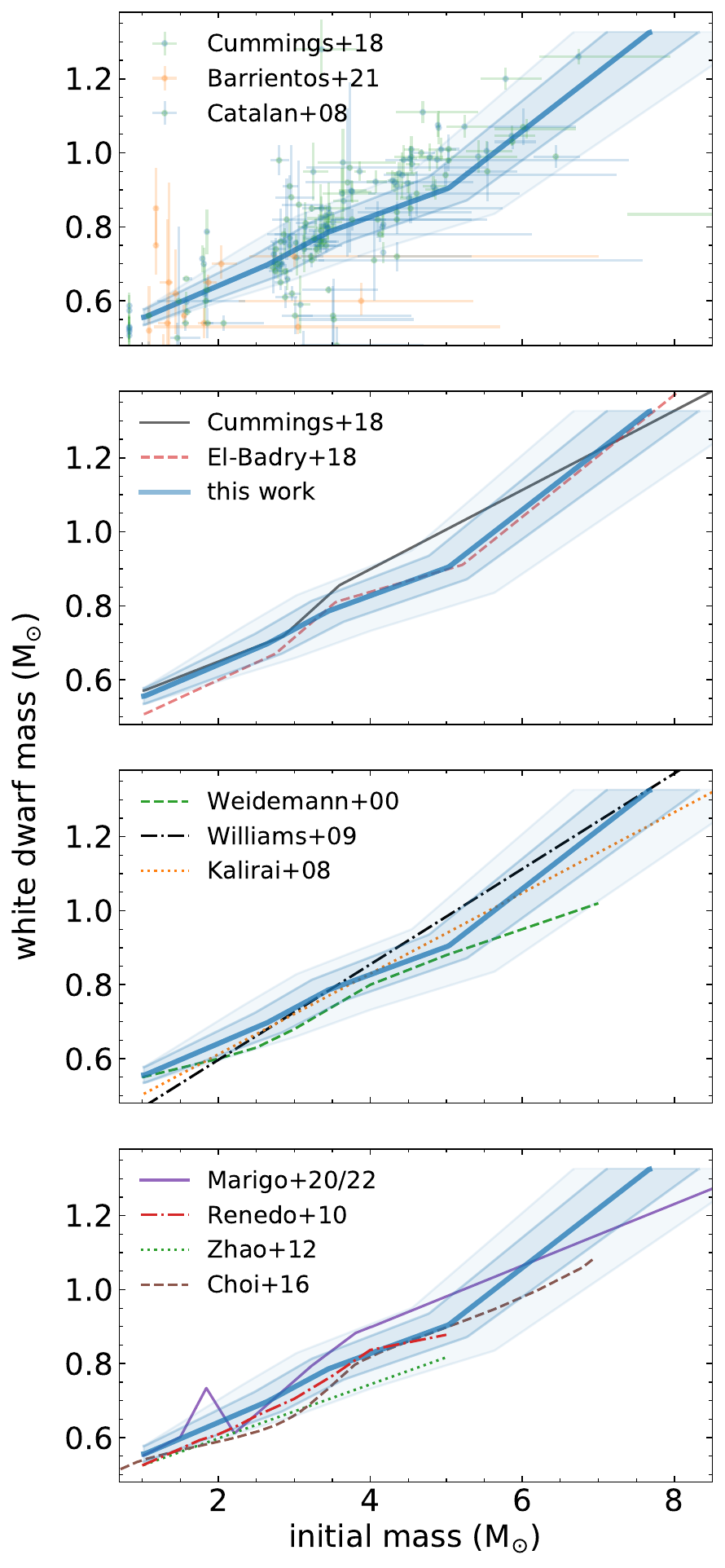}}
	\caption{The initial-final mass relation derived in this work (blue solid line), compared to other prescriptions in the literature. We show the 1- and 2-$\sigma$ uncertainty on the derived IFMR in shaded blue regions. In the top panel we include data from cluster white dwarfs and wide WD+MS binaries from \citet{cummings2018, Barrientos2021,catalan2008} in green, orange and blue error bars, respectively. In the second panel we show the IFMR presented by 
    \citet{cummings2018} which uses clusters and provides the high-mass calibration for this work. We also show the IFMR of \citet{elbadry2018} using the \gaia\ white dwarf sample within 100pc. In the third panel we show other IFMRs derived using clusters, including \citet{weidemann2000,kalirai2008,williams2009}. In the fourth panel we show the theoretical IFMR prescriptions from \citet{renedo2010} and from \citet{choi2016} for Solar abundance. We also show the semi-empirical, non-monotonic IFMR from \citet{marigo2020,Marigo2022} and the IFMR from \citet{zhao2012} which is derived using WD+MS wide binaries.}
	\label{fg:ifmr-1sigma-compare}
\end{figure}

Fig.\,\ref{fg:ifmr-1sigma-compare} shows the 1-$\sigma$ and 2-$\sigma$ uncertainty on the IFMR based on the breakpoints of the piecewise linear fit described above. In this figure the IFMR is plotted by interpolating between the breakpoint locations in Table\,\ref{tb:MC-pwlf-breakpoints}.
The plotted confidence intervals also include the normal distribution of observed measurement uncertainty with standard deviation of 0.02\,\Msun.

We also provide the IFMR in an functional form with the best-fit parameters as follows:

\vspace{2pt}
$(1.0<M_{\rm i} / M_{\odot}<2.5 \pm 0.2)$:
\begin{equation}
M_{\rm WD} = (0.086 \pm 0.003) \times M_{\rm i} + (0.469 \pm 0.004)\,M_{\odot}
\end{equation}

$(2.5 \pm 0.2<M_{\rm i} / M_{\odot}<3.4 \pm 0.1)$: 
\begin{equation}
M_{\rm WD} = (0.10 \pm 0.01) \times M_{\rm i} + (0.40 \pm 0.03)\,M_{\odot} \\ 
\end{equation}

$(3.4 \pm 0.1<M_{\rm i} / M_{\odot}<5.03 \pm 0.08)$: 
\begin{equation}
M_{\rm WD} = (0.06 \pm 0.01) \times M_{\rm i} + (0.57 \pm 0.05)\,M_{\odot} \\ 
\end{equation}

$(5.03 \pm 0.08<M_{\rm i} / M_{\odot}<7.6 \pm 0.3)$:
\begin{equation}
M_{\rm WD} = (0.17 \pm 0.02) \times M_{\rm i} + (0.04 \pm 0.08)\,M_{\odot} 
\end{equation}

The top panel of Fig.\,\ref{fg:ifmr-1sigma-compare} shows the IFMR derived in this study compared with data from three previous studies deriving the IFMR from clusters and wide binaries \citep{cummings2018,catalan2008} and turn-off/subgiant wide binaries \citep{Barrientos2021}. As previously discussed, the IFMR derived here was calibrated against the high-mass IFMR of \citet{cummings2018} (black solid line in second panel of Fig.\,\ref{fg:ifmr-1sigma-compare}). Otherwise, the cluster white dwarf IFMRs are fully independent of this work. The majority of cluster white dwarfs used in previous IFMR prescriptions have had parameters determined using spectroscopy, whereas in this study we exploit \textit{Gaia} photometric parameters. There is also no overlap between the sample used in this study and previous work on clusters since there are no stellar clusters within 40\,pc.

The second panel of Fig.\,\ref{fg:ifmr-1sigma-compare} shows the IFMR derived in this study compared with that of \citet{elbadry2018} and \citet{cummings2018}. The IFMR presented here agrees at the 1-$\sigma$ level with \citet{cummings2018} at most masses, increasing to 2-$\sigma$ agreement in the initial mass range $\approx$3.5--5.5\,\Msun. Despite being calibrated against the \citet{cummings2018} IFMR at high masses, the IFMR derived here closely follows that of \citet{elbadry2018} for $M_{\rm initial}>3.5$\,\Msun, with agreement at the 1-$\sigma$ level. For initial masses $<$2.0\,\Msun, the IFMRs differ by more than 2-$\sigma$ since most of the sample lies in this regime leading to small confidence intervals on both IFMRs.
Our method is very close to that of \citet{elbadry2018}, hence this work provides a validation of their technique. One improvement of the present study is that we did not assume pure-hydrogen atmospheres for \gaia\ photometric mass determinations as done by \citet{elbadry2018}, yet we find a very similar IFMR. This can be understood from their lower temperature limit of 10\,000\,K, above which pure-H, pure-He and mixed model spectra give very similar \gaia\ masses \citep{Gentile2021}. The temperature cut-off used by \citet{elbadry2018} implies that they must fit both the colour and absolute magnitude (i.e. mass and temperature) white dwarf distributions to derive the IMFR. This is a consequence of the mass-dependent cooling rates, resulting in the completeness of their sample being white dwarf mass dependent. Our approach using a volume-limited sample allows us to only consider the one-dimensional white dwarf mass distribution, but the similarity of both IFMRs suggests that the methods are equivalent. The main difference between \citet{elbadry2018} and this work is at the low-mass end, where we have used a different method to constrain the minimum white dwarf mass created from single star evolution in the Galactic disk (see Section\,\ref{sec:PopSyn}).

The third panel of Fig.\,\ref{fg:ifmr-1sigma-compare} also shows the IFMRs derived from three studies utilising cluster white dwarfs \citep{weidemann2000,williams2009,kalirai2008}. We find these to be consistent at the 1-$\sigma$ level with the IFMR presented in this study for initial masses in the range $\approx$2.0--5.5\,\Msun. At larger masses ($>$5.5\,\Msun) the \citet{weidemann2000} IFMR diverges to a 2-$\sigma$ separation, whilst at lower masses ($<2.0$\,\Msun) both the \citet{williams2009} and \citet{kalirai2008} IFMRs exhibit a greater than 2-$\sigma$ separation. 

In the fourth (lowest) panel of Fig.\,\ref{fg:ifmr-1sigma-compare} we show the theoretical IFMRs using different stellar evolution codes \citep{choi2016,renedo2010}, the non-monotonic IFMR from \citet{marigo2020,Marigo2022} and the IFMR from \citet{zhao2012} which is derived using WD+MS binaries. Compared with the present study, we find all four IFMRs to be consistent at the 1--2-$\sigma$ level, with a slight exception for the MIST theoretical IFMR \citep{choi2016} in the initial mass range 2--3\,\Msun, and the semi-empirical non-monotonic peak from the IFMR of \citet{marigo2020} near 2\,\Msun. However, our IMFR derivation method does not allow for non-monotonicity hence we do not rule out that such non-monotonic peak exists.

\section{Conclusions}
\label{sec:conclusions}

In this work have developed a new initial-final mass relation appropriate for the single-star progenitors of white dwarfs. Our method provides a self-consistent determination of the mass of progenitors to the 40\,pc white dwarf sample, which should be of broad utility, especially for studies using \gaia\ derived parameters. The IFMR in this work is broadly consistent with previous studies, finding a 4-piece segmented linear fit to be appropriate. We have accounted for the dominant astrophysical uncertainties around an initial population of main sequence stars, including the gradient of the initial mass function, stellar metallicity and age of the Galactic disc. We have also accounted for the bias that higher-mass white dwarfs are more likely to have been formed from stellar mergers. We have considered two observational samples -- DA-only and the full sample --  in order to explore any bias that may be induced from systematic variations in white dwarf parameter accuracy and spectral evolution. We find that local white dwarfs have liberated between 45--85\% of their initial mass. The mass-loss leading to the white dwarf stage is important as it explains galactic chemical enrichment in the form of gas and dust expelled by stellar winds \citep{Karakas2002,Cristallo2011,ventura18,Ginolfi2018,marigo2020}.

\section*{Acknowledgements}
We are grateful to Karel Temmink for providing the theoretical probabilistic merger fraction data from \citet{temmink2020}.
This research was supported by a Leverhulme Trust Grant (ID RPG-2020-366). Support for this work was provided by NASA through the NASA Hubble Fellowship grant HST-HF2-51527.001-A awarded by the Space Telescope Science Institute, which is operated by the Association of Universities for Research in Astronomy, Inc., for NASA, under contract NAS5-26555. PET and MOB received funding from  from the European Research Council under the European Union’s Horizon 2020 research and innovation programme number 101002408 (MOS100PC).


\section*{Data Availability}
The observational data used in this article are published in \citet{mccleery2020,Gentile2021,OBrien2023} and O'Brien et al., submitted, with the data from the latter available at \url{https://cygnus.astro.warwick.ac.uk/phrtxn/}. The results presented in this study will be made available upon reasonable request to the corresponding author.




\bibliographystyle{mnras}
\bibliography{mybib} 






\begin{figure*}
	\centering
	\subfloat{\includegraphics[width=1.6\columnwidth]{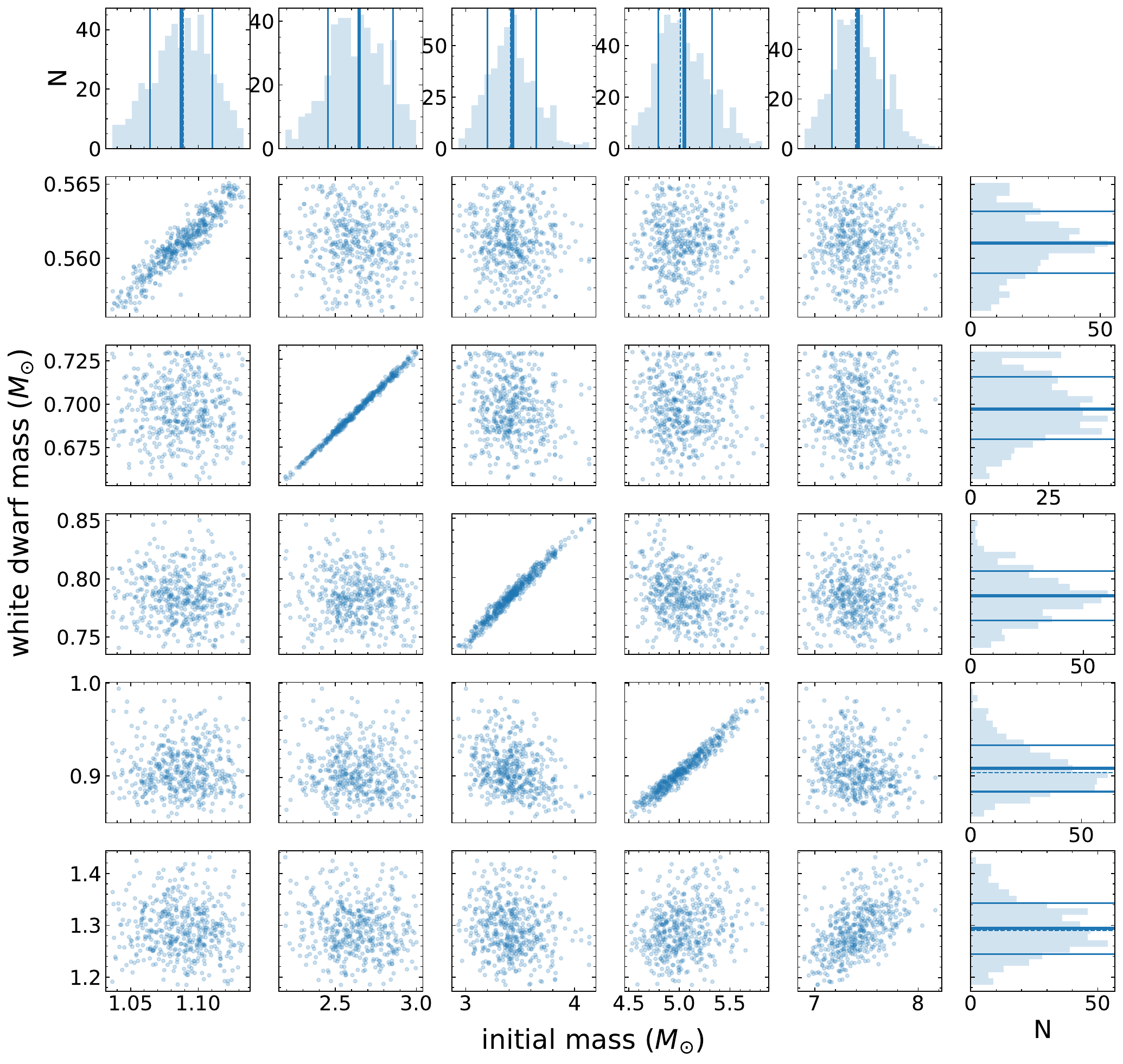}}
	\caption{Best fit break locations in ($M_{\rm in}$, $M_{\rm WD}$) coordinates for the segmented linear regression to 500 Monte Carlo samples of the quantile-quantile IFMR shown in Fig.\,\ref{fg:quantile-pwlf-MC-fits} (and the bottom-right panel of Fig.\,\ref{fg:quantile-param-vary}) which accounts for uncertainties in initial mass function slope, metallicity and Galactic disk age. Top and right panels show distributions in either dimension. Diagonal panels from left-to-right correspond to the breakpoint locations. }
	\label{fg:pwlf-cornerplot}
\end{figure*}


\bsp	
\label{lastpage}
\end{document}